\documentclass[conference]{IEEEtran}

\usepackage{cite}
\usepackage{amsmath,amssymb,amsfonts}
\usepackage{graphicx}
\usepackage{textcomp}
\usepackage{xcolor}
\usepackage{color,soul}
\usepackage{multirow}
\usepackage{url}
\usepackage{caption}
\usepackage{algorithm,algorithmic}
\def\BibTeX{{\rm B\kern-.05em{\sc i\kern-.025em b}\kern-.08em
    T\kern-.1667em\lower.7ex\hbox{E}\kern-.125emX}}

\begin{document}

\title{An MDP-Based Approach for Distribution System Control with PV Generation and Battery Storage}

\author{
    \IEEEauthorblockN{
        Robert Sosnowski\IEEEauthorrefmark{1}\IEEEauthorrefmark{2}, Marcin Baszyński\IEEEauthorrefmark{1}, Charalambos Konstantinou\IEEEauthorrefmark{2}
    }
    \IEEEauthorblockA{\IEEEauthorrefmark{1}Department of Power Electronics and
    Energy Control Systems, AGH University of Science and Technology (AGH UST)}
    \IEEEauthorblockA{\IEEEauthorrefmark{2}CEMSE Division, King Abdullah University of Science and Technology (KAUST)}
    \IEEEauthorblockA{E-mail: \{sosnowski, mbaszyn\}@agh.edu.pl, \{robert.sosnowski, charalambos.konstantinou\}@kaust.edu.sa}
}

\maketitle

\begin{abstract}
%The photovoltaic (PV) units integration in the distribution system is an important problem that requires the development of proper control approaches.
This paper proposes a decision-making approach for the control of distribution systems with distributed energy resources (DERs) equipped with photovoltaic (PV) units and battery energy storage systems (BESS). The objective is to minimize the total operational cost of the distribution system while satisfying the system operating constraints. The method is based on the discrete-time finite-horizon Markov Decision Process (MDP) framework. Different aspects of the distribution system operation are considered, such as the possibilities of curtailment of PV generation, managing battery storage, reactive power injection, load shedding, and providing a flexibility service for the transmission system. The model is tested for the IEEE 33-bus system with two added DERs and the study cases involve various unexpected events. The experimental results show that this method enables the attainment of relatively low total cost values compared to the reference deterministic approach. The benefits of applying this approach are particularly evident when there is a significant difference between the predicted and actual PV power generation.
\end{abstract}
%\vspace{-3mm}
%\vspace{-3mm}
%\section*{Nomenclature}
%\addcontentsline{toc}{section}{Nomenclature}
%\begin{IEEEdescription}[\IEEEusemathlabelsep\IEEEsetlabelwidth{$\Delta P, \Delta P_{Load}$}]
%\item[$a, s$] Action, state
%\item[$\mathbb{A},\mathbb{A}_{on}, \mathbb{S}$] Offline and online actions sets, states set
%\item[$b/d/n/l$] BESS/DER/bus/line index
%\item[$\mathbb{B}/\mathbb{D}/\mathbb{N}/\mathbb{L}$] BESS/DER/bus/line set
%\item[$c, C$] Price, cost
%\item[$E_{bes,b}$] Energy stored in BESS $b$
%\item[$P/Q/|S|$] Active/reactive/apparent power
%\item[$P_{pv}^{pred},\delta P_{pv}$] Prediction and its relative error of the PV power generation
%\item[$\Delta P, \Delta P_{Load}$] Power losses, load shedding power
%\item[$T^{a}(s'|s)$] Transition probability from state $s$ to $s'$ under action $a$
%\item[$t/\Delta t/\mathbb{T}$] Time epoch number/duration/set
%\item[$U_{t}(s)$] Expected utility for epoch $t$ and state $s$
%\item[$V_n, I_l$] Bus $n$ voltage, line $l$ current (RMS)
%\item[$\gamma$] Discounting factor
%\end{IEEEdescription}
%\vspace{-2mm}
\section{Introduction}
The reduction of operational costs, alongside with the increasing load and the need to reduce emissions to the environment, requires the development of optimal energy management algorithms for distribution systems~\cite{Alam}.
One of the major challenges regarding modern electric grids is the integration of renewable energy sources (RES) such as photovoltaic (PV) systems. Despite that distributed RES generation decreases net power demand and the power losses, the relatively high uncertainty of such generation may affect the grid reliability by causing some additional challenges such as voltage fluctuations \cite{Judge}. 
Furthermore, RES can be affected by extreme weather conditions, e.g., lightning, wildfire, flood, snow or dust cover~\cite{Bosnjakovic}, and operational faults, e.g., line to ground or inverter faults \cite{Osmani}, resulting in reduced or interrupted generation.

Energy storage systems (ESS) can play an important role in dealing with the stochastic nature of renewable energy resources~\cite{Calero}. Specifically, battery energy storage  systems (BESS), which are characterized by very fast response and capability of providing regulation services~\cite{Calero}, are becoming more and more popular in recent years, e.g., the United States battery storage capacity is expected to grow by 89\% in 2024 \cite{Antonio}. ESS can help in dealing with the intermittent generation and fluctuations of the power generated by renewable resources and enhance system's efficiency and flexibility~\cite{Tan}.
% MDP

Considering the intermittency of DER-based RES as well as the integration of BESS in distribution systems, it is of paramount importance to develop decision-making methods that are able to optimize the system operation and reduce operational costs. 
%for the distribution system with distributed energy resources (DERs) equipped with PV panels and battery storage, which enable to deal with the generation uncertainty related to the normal weather changes, and the consequences of the faults and extreme events.
One of the frameworks which can be used for modeling systems with uncertainty is the Markov Decision Process (MDP). 
%\subsection{Related work}
In literature, there are many works regarding approaches which are based on MDP or Markov Chain which focus on thermostatically controlled loads (TCLs) control in order to provide ancillary services for the electric grid. 
For instance, such works focus on TCL synchronization \cite{Angeli, Meyn}, lack of real-time state information \cite{Mathieu}, demand response to compensate the forecast error of wind power generation \cite{Z.Li}, and taking into account both utility, aggregator and customer perspectives \cite{Hassan, Hassan2}.

MDP-based methods can also be applied for managing the power system with wind generators by controlling the energy storage \cite{Zephyr}, or system configuration  \cite{Wang}. These papers present methods for reducing the calculation time by using stochastic dual dynamic programming (SDDP) and approximate dynamic programming (ADP) approaches, respectively. MDPs can be also applied to solve unit commitment and economic dispatch problems  with intermittent wind generation \cite{Luh,Yu}. In the area of distribution systems with PV generation, the authors in \cite{Vergara} present an approach based on reinforcement learning to optimally dispatch PV inverters in unbalanced distribution systems using a decentralized architecture. In, \cite{Grillo}, a distribution system with both PV and ESS is considered in which the operator can control the energy storage by specifying the amount of energy that is drawn from or supplied to it, in order to minimize the cost of the network power losses. MDP-based approaches are also applicable to deal with the consequences of extreme weather conditions.
For instance, in \cite{Wang2}, the authors present an approach applicable for unfolding events such as typhoons. Another example is optimal load restorations in distribution systems by dispatching distributed generation and ESS units after sudden outages caused by, e.g., extreme natural events such as hurricanes \cite{Hosseini}. 

%\subsection{Contribution and paper structure}
While past works concentrate on a relatively narrow scope of the operational aspects of the distribution system, this work takes into account different aspects of the distribution system operation, such as the possibilities of PV generation curtailment, management of BESS, reactive power injection, load shedding, and providing a flexible service to the transmission system. It presents a decision-making method using the MDP discrete-time finite-horizon framework for the distribution system control.  
The proposed approach, in relation to the PV generation probabilities, takes advantage of the PV generation predictions from the previous day by considering relative errors of these predictions as MDP model states, and enables to use the same probability matrix for each time of the day and year. 
The presented method can be applied not only as a decision-making tool, but also as an analysis tool for the distribution system operator (DSO). It can be used to design the components of the distribution system, for example to decide on the placement and sizing of DERs, based on the expected total cost value for specific cases. 
Various study cases are considered involving unexpected events, such as temporary shut-down of PV generation, significant difference between the prediction and actual values of PV units, unpredicted load increase, and unplanned flexibility service demand. The obtained results are compared with two reference models.

The reminder of the paper is as follows. In Section \ref{s:proposed_method} the proposed formulation is described. Section \ref{s:model} contains the parameters of the system and the MDP model used for the calculations as well as the descriptions of the considered study cases. Section \ref{s:results} presents the obtained results for the proposed and reference methods. Section \ref{s:conclusion} concludes the work.
%\vspace{-2mm}
\section{Proposed MDP-Based Formulation}\label{s:proposed_method}

In this section, we present the formulation of the proposed MDP-based decision-making method for the control of the distribution grid with DERs. This method is based on the discrete-time finite-horizon MDP, and its objective is to minimize the expected operational cost incurred by the DSO. {The nomenclature used in the work is provided in Table~\ref{t:nomenclature}.}

\begin{table}[t]
    \centering
    \vspace{3pt}
    \caption{Nomenclature.}
    \begin{tabular}{*{2}{c}}
    \hline
    Symbol & Description\\
    \hline
    $a, s$ & Action, state\\
    $\mathbb{A},\mathbb{A}_{on}, \mathbb{S}$ & Offline and online actions sets, states set\\
    $b/d/n/l$ & BESS/DER/bus/line index\\
    $\mathbb{B}/\mathbb{D}/\mathbb{N}/\mathbb{L}$ & BESS/DER/bus/line set\\
    $c, C$ & Price, cost\\
    $E_{bes,b}$ & Energy stored in BESS $b$\\
    $P/Q/|S|$ & Active/reactive/apparent power\\
    $P_{pv}^{pred},\delta P_{pv}$ & Prediction and its relative error of the PV power gen.\\
    $T^{a}(s'|s)$ & Transition probability from state $s$ to $s'$ under action $a$\\
    $t/\Delta t/\mathbb{T}$ & Time epoch number/duration/set\\
    $U_{t}(s)$ & Expected utility for epoch $t$ and state $s$\\
    $V_n, I_l$ & Bus $n$ voltage, line $l$ current (RMS)\\
    $\gamma$ & Discounting factor\\
    \hline
    \end{tabular}
    %\vspace{-4mm}
    \label{t:nomenclature}
\end{table}\par

\subsection{System Constraints}
The optimal solution is determined considering the system constraints. These are distribution system constraints, i.e., bus voltage limits \eqref{A}, line current limits \eqref{B}, and DERs limits such as inverter apparent power \eqref{C}, battery capacity \eqref{D} and battery's converter power limits \eqref{E}.
\begin{subequations}
\begin{align}
    V_{n}^{min} \leq V_{n,t} \leq V_{n}^{max} \ \ \ \forall n \in \mathbb{N}, \forall t \in \mathbb{T} \label{A} \\
    I_{l,t} \leq I_{l}^{max}  \ \ \ \forall l \in \mathbb{L}, \forall t \in \mathbb{T} \label{B} \\
    |S|_{inv,d,t} \leq |S|_{inv,d}^{max}  \ \ \ \forall d \in \mathbb{D}, \forall  t \in \mathbb{T} \label{C} \\
    E_{bes,b}^{min} \leq E_{bes,b,t} \leq E_{bes,b}^{max}  \ \ \ \forall b \in \mathbb{B}, \forall t \in \mathbb{T} \label{D} \\
    P_{bes,b}^{min} \leq P_{bes,b,t} \leq P_{bes,b}^{max}  \ \ \ \forall b \in \mathbb{B}, \forall t \in \mathbb{T} \label{E}
\end{align}\par
In addition, there are power transfer limits of the transmission system, which are defined by the transmission system operator (TSO). These limits could be time-varying during the day and represented by P\&Q charts, so the limits of the active power $P_{TS,t}$ for the time epoch $t$ depend on the value of the reactive power $Q_{TS,t}$ and vice versa. It is assumed that these limits are defined by TSO in such way that the apparent power limit for the systems' point of connection is followed:
\begin{align}
    |S|_{TS,t} \leq |S|_{TS}^{max} \ \ \ \forall t \in \mathbb{T} \label{F}
\end{align}
\end{subequations}\par

\subsection{MDP Model}
\subsubsection{States}
The states we consider in the MDP formulation are the following: the relative error of the PV generation prediction ($s_{pv}$) for each DER, and the battery storage level ($s_{bes}$) for each BESS. The state can be represented as a vector:
\begin{equation}
    s=[s_{pv,1};...;s_{pv,D};s_{bes,1};...;s_{bes,B}]
\end{equation}
where $D$ is the number of DERs and $B$ is number of the BESS  units. 
The state space is discretized by dividing values into intervals. 
The relative error of the PV power generation prediction for each DER is independent of actions and it is considered as stochastic. In order to limit the values in the range from $-1$ to $1$, this relative error is defined as follows:
\begin{equation}
    \delta\\P_{pv,t}=\frac{P_{pv,t}-P_{pv,t}^{pred}}{\max{\left\{P_{pv,t}; P_{pv,t}^{pred}\right\}}}
\end{equation} 
The battery storage level is expressed as a relatively value in range 0 to 100\%.
The load demand is not considered as a state as it is independent from the actions and can be assumed as deterministic due to the fact that load forecasts are relatively accurate, e.g., the mean absolute error of day-ahead forecast is 1-3\% of the load \cite{Luh}.

\subsubsection{Actions}
The DSO is assumed to decide on the PV generation curtailment ($a_{pv}$) and the injection of reactive power ($a_{q}$) for each DER, charging and discharging the DERs batteries ($a_{bes}$) for each BESS, and load shedding ($a_{load}$). The actions can be represented as a vector:
\begin{equation}
    a=[a_{pv,1};a_{q,1};...;a_{pv,D};a_{q,D};a_{bes,1};...;a_{bes,B};a_{load}]
\end{equation} 
For each value of $a_{load}$ there are assigned numbers of buses which are affected by load shedding. Other types of actions are expressed as a relative value and divided into equal intervals. 

\subsubsection{Transition Probabilities}
The battery storage transitions are assumed to be deterministic, thus, $T_{bes}^{a_{bes}}(s_{bes}'|s_{bes})$, which is the probability %\textcolor{blue}{
of the transition from the BESS state $s_{bes}$ to state $s'_{bes}$ under the BESS action $a_{bes}$, 
%}
is equal $0$ or $1$. For night hours, the PV generation is equal zero, so the model is fully deterministic during this time. A stochastic approach is only applied to the PV generation for daylight hours. 
The transition probabilities from a particular PV state ($s_{pv}$) to the another ($s_{pv}'$) are calculated based on historical data as a ratio of the occurrences of this transition ($N_{tran,s_{pv},s_{pv}'}$) to the total number of occurrences of that particular state ($N_{occur,s_{pv}}$):
\begin{equation}
    T_{pv}(s_{pv}'|s_{pv})=\frac{N_{tran,s_{pv},s_{pv}'}}{N_{occur,s_{pv}}} 
\end{equation}
The transition probabilities for each PV unit $d$ ($T_{pv,d}$) and BESS $b$ ($T_{bes,b}$) are independent from others, hence, the resultant transition probability is equal to the product of the all probabilities. Accordingly, the resultant transition probability from particular state $s$ to state $s'$ under action $a$ is as follows:
\begin{equation}
    T^{a}(s'|s)=\prod_{\substack{b\epsilon \mathbb{B}, \\ d\epsilon \mathbb{D}}} T_{bes,b}^{a_{bes,b}}(s_{bes,b}'|s_{bes,b}) \cdot T_{pv,d}(s_{pv,d}'|s_{pv,d})  
\end{equation}

\subsubsection{Costs}
Different types of costs are considered in our MDP formulation: penalty for the curtailment of PV generation representing equivalent cost for the under-utilization of the PV resources ($C_{PV}$), penalty for load shedding ($C_{Load}$), and equivalent cost of power losses for both the grid ($C_{Grid}$) and DERs ($C_{DER}$).
These costs are calculated as a product of the energy price (which can be defined separately for each type of costs and each time epoch), the value of power curtailment ($\Delta P_{pv}$), load shedding ($\Delta P_{Load}$) or losses ($\Delta P_{Grid}$ and $\Delta P_{DER}$), and duration of the time epoch. 
Additionally, we consider the reward from the TSO for providing the flexibility service ($C_{TSO}$) ($C_{TSO} \leq 0$). This reward is defined by the TSO based on the active ($P_{TS}$) and reactive ($Q_{TS}$) power transfer between the distribution and transmission system. Thus, the total cost for a specific time epoch $t$ is equal to the sum of mentioned values:
\begin{equation}
\begin{split}
C_{t}= \Delta t \cdot (c_{PV,t} \cdot \Delta P_{PV,t} + c_{Load,t} \cdot \Delta P_{Load,t} \\ 
+ c_{Grid,t} \cdot \Delta P_{Grid,t} + c_{DER,t} \cdot \Delta P_{DER,t}) \\ 
+ C_{TSO,t}(P_{TS,t},Q_{TS,t})
\end{split}
\end{equation}
%where: $\alpha=\{PV, Load, Grid, DER\}$.

\subsection{Expected Utility}
\begin{figure}[t]
    \centering
    \includegraphics[trim=0cm 0cm 0cm 0cm, clip, width=3.4in]{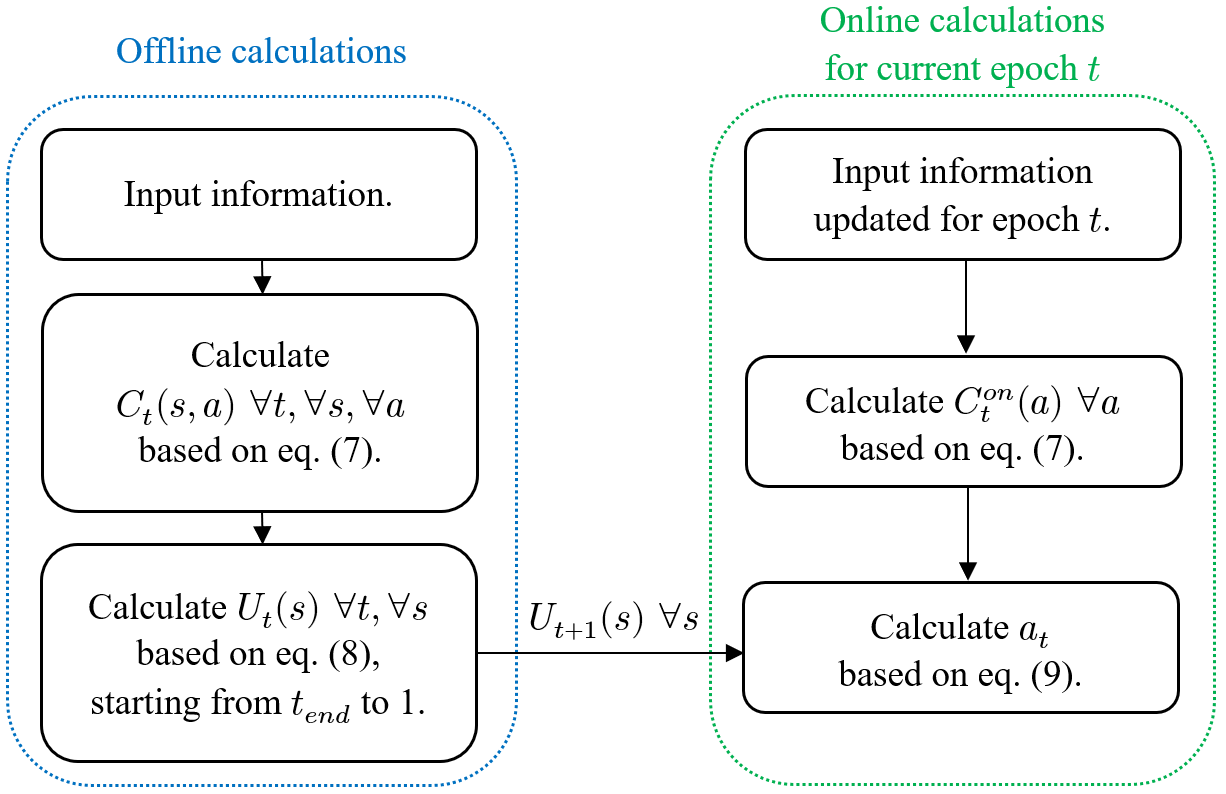}
    \caption{{General flow chart of the offline and online calculations.}}
    %\vspace{-3mm}
    \label{fig:flowchart}
\end{figure}

The calculations of the expected utility (total cost) are divided into two stages, offline and online. A flow chart of the calculations is presented in Fig.~\ref{fig:flowchart}. 

\subsubsection{Offline Calculations}
The offline calculations are performed the day before. The result of these calculations is the expected utility value for each epoch and state. It is assumed that the information known one day ahead is the following: the load demand predictions for each bus, the PV generation predictions for each DER, and planned by TSO: the power transfer constraints and the flexibility service demand with appropriate information about the reward for providing the service. 
For each time epoch, state, and action, the value of cost $C_t(s,a)$ is calculated and the constraints are checked. In case of constraints violation, the cost value is replaced by a very large number. Then the expected utility is calculated for each time epoch and state using the backward induction method:
\begin{equation}
    U_{t}(s) = \min_{a \in \mathbb{A}} [C_t(s,a) + \gamma^{t}  \sum_{s'} T_{t}^{a}(s'|s) \cdot U_{t+1}(s')] \ 
\end{equation} 

\subsubsection{Online Calculations}
The offline calculations will lead inevitably to errors due to  the inaccurate forecasting of the load demand, planned flexibility value provided by the TSO, and the quantized values of the power generation. In order to decrease an impact of these inaccurate estimations, the immediate cost $C_t^{on}$ for each action is recalculated using the actual data.
It is assumed that for each time epoch in real-time, the DSO is provided with the actual values of the load demand, PV generation as well as information about transmission system constraints and flexibility service demand. 
Additionally, since all actions except $a_{bes}$ have no impact on the probabilities of state transitions, it is possible to extend the range of actions ($\mathbb{A}_{on}$) for the current epoch without the need to modify $T^{a}(s'|s)$ or $U_{t+1}(s')$. The selected action minimizes the expected cost value.
\begin{equation}
    a_{t}=\arg\min_{a \in \mathbb{A}_{on}} [C_t^{on}(a) + \gamma^{t} \sum_{s'} T_{t}^{a}(s'|s_{t}) \cdot U_{t+1}(s')]
\end{equation}
\section{Experimental Setup, Parameters, and Cases}\label{s:model}
\subsection{Grid and DERs Parameters}
\begin{figure}[t]
    \centering
    \includegraphics[trim=1.5cm 6cm 1.5cm 6.5cm, clip, width=3.4in]{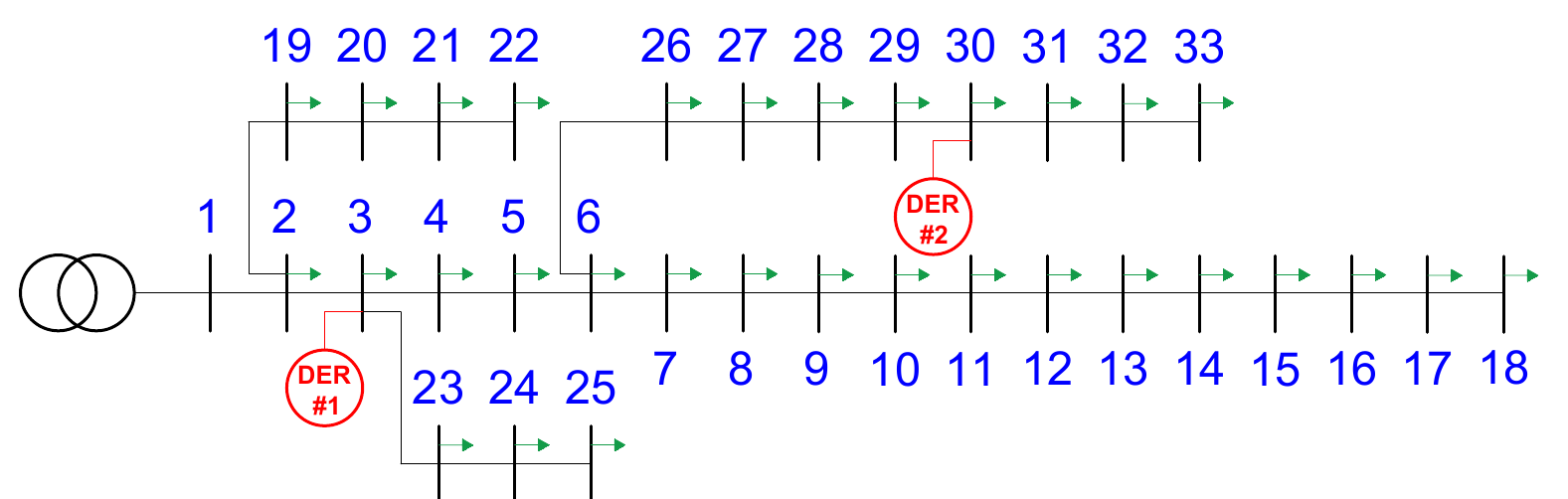}
    \caption{{{Single-line diagram of the IEEE 33-bus distribution system with added DERs.}}}
    %\vspace{-1mm}
    \label{fig:grid}
\end{figure}

%\subsubsection{Grid}
The IEEE 33-bus distribution system is considered with default parameters and topology as {presented in Fig.~\ref{fig:grid}} \cite{Baran}. The slack bus voltage is set to $1.05$ p.u. and its maximum apparent power equals $4$~MVA.
The lower and upper voltage limits for all buses are $0.95$ p.u. and $1.05$ p.u., respectively, and the line current limits are  $200$ A for lines 1-2, $150$ A for lines 3-5, $100$ A for lines 6-7 and 22-29, and $50$ A for other lines. 
The power flow is calculated using the backward forward sweep method \cite{Rupa}, and only resistive losses are considered for the lines. 

%\subsubsection{DERs}
According to \cite{Shijie}, the optimal placement and sizing of two DERs in terms of minimizing power losses are buses 3 and 30 with power injections equal $2.55$~MW and $1.16$~MW, respectively. Assuming the inverter efficiency (including the power losses of the transformer and interconnection lines) is equal 90\%, the inverter power ratings are selected with some margin as $3$~MVA and $1.5$~MVA, respectively. The first DER is equipped with PV panels and battery storage (each $3$ MW of the power rating) with common DC link. The battery capacity ($2.85$ MWh) is selected in such a way that it is possible to fully charge the battery in 1 hour, assuming the efficiency of the battery converter is equal to 95\%. The second is equipped only with a PV unit ($1.5$~MW). 

\subsection{MDP Model Parameters}
\begin{figure}[t]
    \centering
    \includegraphics[trim=0.5cm 0cm 0.5cm 0cm, clip, width=3.4in]{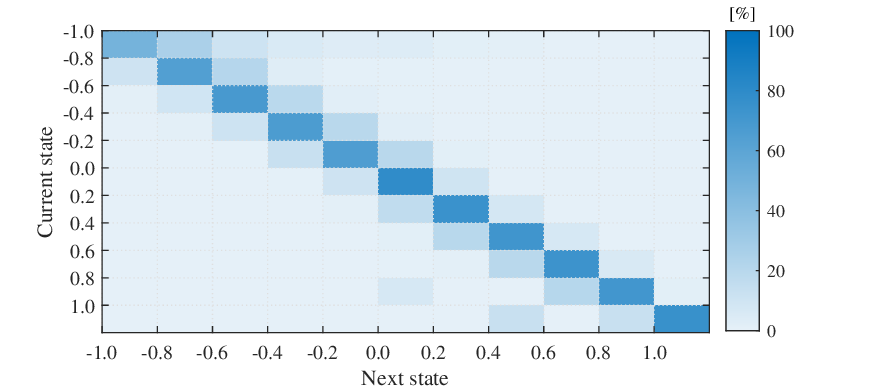}
    \caption{{Transition probability matrix for the relative error of the PV prediction.}}
    %\vspace{-5mm}
    \label{fig:transition}
\end{figure}

The time duration of the epoch is 15 minutes and the calculation horizon is one day (24 hours, 96 epochs). The discounting factor $\gamma$ is equal to $1$.  Table \ref{t:statesactions} presents the ranges of values (minimum and maximum values), the intervals between them, and the number of values within each range. The values of $s_{pv}$, $a_{pv}$ and $a_{q}$ are defined separately for both DERs. The possibility of extending the set of actions for online calculations is demonstrated using the example of reactive power actions (an extension from 5 to 9 values), for which the offline calculation interval is the greatest. The load shedding is performed by disconnecting the load for a chosen group of buses: 2-33 (all buses), 2-28, 19-33, 2-10, 11-18, 19-25, 26-33, 2-6, 7-10, 11-14, 15-18, 19-22, 23-25, 26-29, 30-33 or none of the buses. 
The transition probability matrix (Fig. \ref{fig:transition}) for PV states is determined based on rescaled historical data~\cite{Elia}.

\begin{table}[t]
    \centering
    \vspace{3pt}
    \caption{States and actions values.}
    \begin{tabular}{*{5}{c}}
    \hline
    state/action & min. [\%] & max. [\%] & interval [\%] & number\\
    \hline
    $s_{pv1}$, $s_{pv2}$ & -100 & 100 & 20 & 11\\
    %\hline
    $s_{bes}$ & 0 & 100 & 6.25 & 17\\
    %\hline
    $a_{pv1}$, $a_{pv2}$ & 0 & 100 & 25 & 5\\
    %\hline
    $a_{bes}$ & -25 & 25 & 6.25 & 9\\
    %\hline
    $a_{q1}$, $a_{q2}$ & -100 & 100 & 50 / 25 & 5 / 9\\
    %\hline
    $a_{load}$ & none & all buses & - & 16\\
    \hline
    \end{tabular}
    %\vspace{-4mm}
    \label{t:statesactions}
\end{table}\par
%To total number of states is equal 2057, the number of actions is 90000 for online calculations, and 291600 for online calculations.\par

%\subsubsection{Transition probabilities}

%\vspace{-2mm}
\subsection{Power Profiles and Costs}
\subsubsection{Load Demand}
\begin{figure}[t]
    \centering
    \includegraphics[trim=0.5cm 0cm 0.5cm 0cm, clip, width=3.4in]{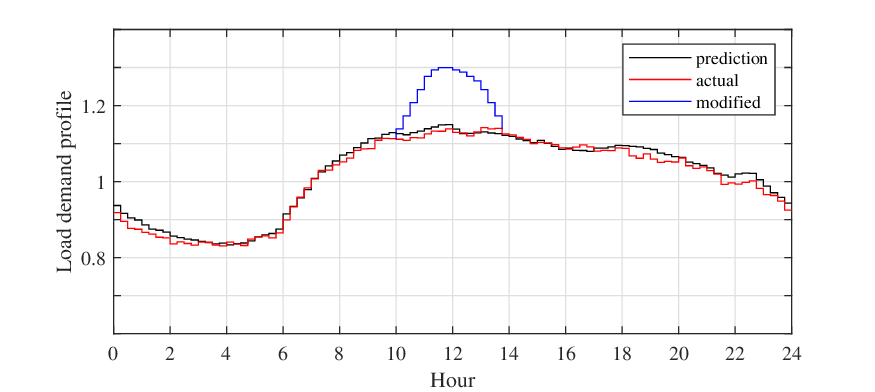}
    \caption{{Load demand profiles (relative to the nominal): predicted, actual, and modified (by adding an unpredicted increase between 10:00 and 14:00).}}
    %\vspace{-3mm}
    \label{fig:load}
\end{figure}

The load demand and its prediction for all buses is equal to the multiplication of the nominal load and the load profile factor, which is assumed to be the same for all buses~\cite{Baran}. 
The values of this factor are defined based on the rescaled historical data~\cite{Elia}. Furthermore, a modification of the actual load demand profile is prepared by adding of an unpredicted load increase between 10:00 and 14:00 (Fig.~\ref{fig:load}).

\subsubsection{PV Generation}
\begin{figure}[t]
    \centering
    %\vspace{-2mm}
    \includegraphics[trim=0.5cm 0cm 0.5cm 0cm, clip, width=3.4in]{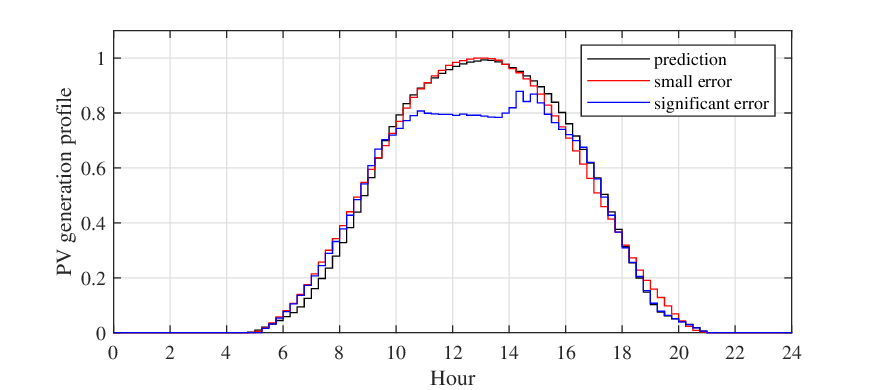}
    \caption{{PV generation profiles (relative to the maximum) for the first DER: predicted and actual values with small and significant error.}}
    \label{fig:pv1}
    %\vspace{-4mm}
\end{figure}

\begin{figure}[t]
    \centering
    \includegraphics[trim=0.5cm 0cm 0.5cm 0cm, clip, width=3.4in]{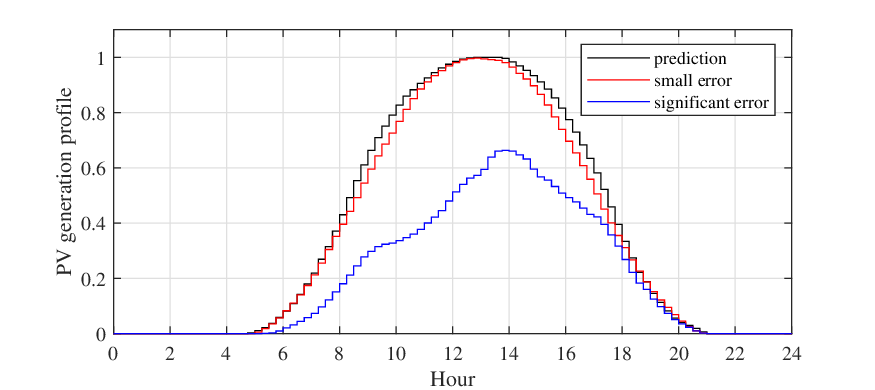}
    \caption{{PV generation profiles (relative to the maximum) for the second DER: predicted and actual values with small and significant error.}}
    %\vspace{-4mm}
    \label{fig:pv2}
\end{figure}

The PV generation profiles, both actual values and predictions, are also determined based on the historical data \cite{Elia}. Two variants of actual values are considered: with relatively small error of the generation prediction and with significant errors, i.e., significantly lower generation than prediction for both DERs (Fig. \ref{fig:pv1} and Fig. \ref{fig:pv2}).

\subsubsection{TSO Constraints and Flexibility Service Demand}
\begin{figure}[t]
    \centering
    \includegraphics[trim=0.5cm 0cm 0.5cm 0cm, clip, width=3.4in]{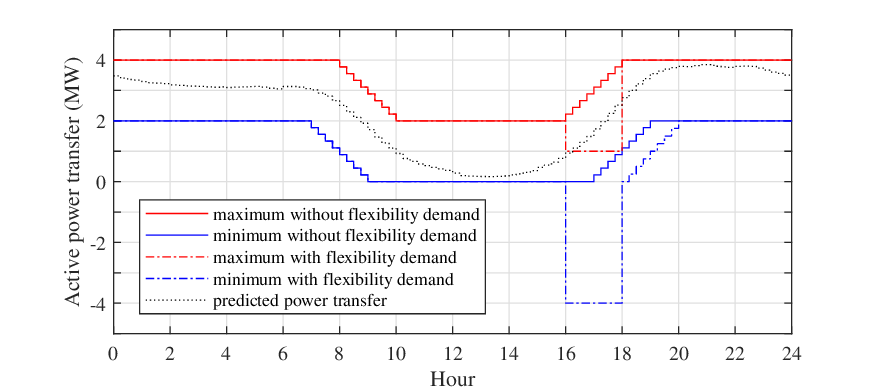}
    \caption{{The limits of the active power transfer between the transmission and distribution system in two variants: with and without flexibility service demand.}}
    %\vspace{-3mm}
    \label{fig:tso}
\end{figure}

Two variants of the TSO power transfer limits are considered: without and with flexibility service demand, presented in Fig.~\ref{fig:tso}.
These limits are determined based on the predicted power transfer (equal to the difference between predictions of the load demand and PV units injections).

\subsubsection{Costs}
The energy price is assumed to be constant during the day and equal 200~\$/MWh for all types of power losses and PV generation curtailment.
The penalty cost for load shedding is assumed to be 600~\$/MWh. 
It is assumed that the transmission system requires active power contribution between 16:00 and 18:00 and the flexibility service reward for providing the power transfer below the upper limit ($P_{TS,t}^{max}=1$ MW) is equal to $100$~\$/MW for each epoch.

%\vspace{-2mm}
\subsection{Case Studies}
We consider five study cases in our experimental setup and evaluation process:
\subsubsection{Default}
The actual load demand (without modifications) and PV generation profiles with small predictions errors are used. The flexibility service is planned, so it is taken into account in the offline calculations.
\subsubsection{Temporary Shut Down of PV Generation (DER \#2)} 
The difference between this study case and the default one is the modification of the PV generation profile of the second DER in such a way that there is an unplanned shut-down (zero generation) between 12:00 and 14:00.
\subsubsection{Significantly Lower PV Generation than the Predicted}
This study case differs from the default one by using the PV generation profiles with significantly lower generation than the prediction.
\subsubsection{Lower PV Generation and Unpredicted Load Demand Increase}
Compared to the previous study case, there is additionally a modified load demand profile with an unpredicted load increase in hours 10:00-14:00.
\subsubsection{Lower PV Generation and Unplanned Flexibility Service Demand}
In this case, in addition to the lower PV generation, there is an unplanned flexibility service demand. Hence, in the offline calculations there are used TSO constraints values without flexibility demand modifications and there is no expected TSO reward. However, in the online calculations, the flexibility service demand and reward are exactly the same as in the previous cases. 
\section{Simulation Results}\label{s:results}
\subsection{Reference Models}
In order to perform a comparison of the results obtained for the proposed method, two reference models are defined. 
The set of actions, constraints, and parameters for the reference approaches is assumed to be exactly the same as for the proposed method.
\subsubsection{Deterministic Model}
The main property of the proposed MDP model is the stochastic approach to PV generation, hence, in order to present the potential benefits of applying this approach, as a first reference method we consider a deterministic model. 
The computational burden for the deterministic method is much lower than for the proposed method, therefore, it is possible to recalculate the problem for each time epoch using the most recent available predictions.
In order to update the predictions for each epoch, the modified smart persistence model is used \cite{Iheanetu}. This model is based on the assumption that the current ratio between the actual generation and the predicted one for a clear sky remains the same for the following epochs. The introduced modification involves replacing the clear sky predictions with the standard predictions from the day before. The prediction for epoch $t'$ updated in epoch $t$ is calculated as follows:
\begin{equation}
    P_{pv,t'}^{update,t}=
    \begin{cases}
        P_{pv,t'}^{pred} & \text{if } P_{pv,t}^{pred} = 0\\
        P_{pv,t'}^{pred} \cdot \frac{P_{pv,t}}{P_{pv,t}^{pred}} & \text{otherwise} 
    \end{cases}
\end{equation}

\subsubsection{Optimal Decisions Model}
As the second reference approach, an optimal decisions model is used.
For this model, it is assumed that the actual PV generation profiles are known at start of the day, the minimal cost value which could be obtained by choosing the optimal action for each epoch is determined. This value is the lower bound of the cost which is not possible to exceed for the considered assumptions.

\subsection{Discussion of Obtained Results}
The time of the offline and online calculations (one epoch) is equal to approx. 11~hours and 25 seconds, respectively, using a computing system with 64-core 3.6~GHz processor and 1.5~TB~RAM.
{The offline computation time is significantly prolonged due to the cost calculations, including power flow, for every epoch ($96$), state ($11^2\cdot17=2057$), and action ($5^2\cdot9\cdot5^2\cdot16=90000$), where the number of combinations is enormous ($96\cdot2057\cdot90000=1.78\cdot10^{10}$). 
Table \ref{t:costs} presents the total cost and its distribution between types for all considered cases and models, and the excessive cost value (over the optimal value for considered assumptions).
Additionally, the total cost comparison is presented in the form of a bar plot in Fig. \ref{fig:costs}. 

\begin{figure}[t]
    \centering
    \includegraphics[trim=0.5cm 0cm 0.5cm 0cm, clip, width=3.4in]{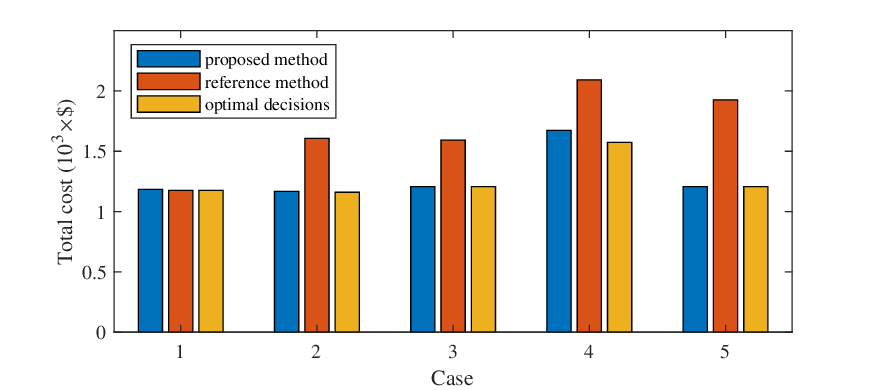}
    \caption{{The total cost comparison for each considered case study and method.}}
    %\vspace{-3mm}
    \label{fig:costs}
\end{figure}\par

\begin{table}[t]
    \centering
    \vspace{4pt}
    \caption{Costs comparison [$10^3\times $\$].}
    \begin{tabular}{*{5}{c}}
    \hline
        \multirow{2}{*}{Case} & \multirow{2}{*}{Cost} & Proposed & Reference & Optimal\\
        &&method & method & decisions\\
    \hline
         & $C_{Grid}$ & \phantom{$-$}0.566 & \phantom{$-$}0.562 & \phantom{$-$}0.562\\
         & $C_{DER}$ & \phantom{$-$}0.809 & \phantom{$-$}0.805 & \phantom{$-$}0.805\\
         & $C_{PV}$ & \phantom{$-$}0.000 & \phantom{$-$}0.000 & \phantom{$-$}0.000\\
        1 & $C_{Load}$ & \phantom{$-$}0.044 & \phantom{$-$}0.044 & \phantom{$-$}0.044\\
         & $C_{TSO}$ & $-$0.236 & $-$0.236 & $-$0.236\\
         & Total & \phantom{$-$}1.184 & \phantom{$-$}1.175 & \phantom{$-$}1.175\\
         & Excessive & \phantom{$-$}0.009 & \phantom{$-$}0.000 & \phantom{$-$}-\\
    \hline
         & $C_{Grid}$ & \phantom{$-$}0.610 & \phantom{$-$}0.599 & \phantom{$-$}0.606\\
         & $C_{DER}$ & \phantom{$-$}0.749 & \phantom{$-$}0.748 & \phantom{$-$}0.745\\
         & $C_{PV}$ & \phantom{$-$}0.000 & \phantom{$-$}0.000 & \phantom{$-$}0.000\\
        2 & $C_{Load}$ & \phantom{$-$}0.044 & \phantom{$-$}0.495 & \phantom{$-$}0.044\\
         & $C_{TSO}$ & $-$0.236 & $-$0.236 & $-$0.236\\
         & Total & \phantom{$-$}1.168 & \phantom{$-$}1.606 & \phantom{$-$}1.160\\
         & Excessive & \phantom{$-$}0.008 & \phantom{$-$}0.446 & \phantom{$-$}-\\
    \hline
         & $C_{Grid}$ & \phantom{$-$}0.619 & \phantom{$-$}0.608 & \phantom{$-$}0.619\\
         & $C_{DER}$ & \phantom{$-$}0.640 & \phantom{$-$}0.639 & \phantom{$-$}0.640\\
         & $C_{PV}$ & \phantom{$-$}0.000 & \phantom{$-$}0.000 & \phantom{$-$}0.000\\
        3 & $C_{Load}$ & \phantom{$-$}0.132 & \phantom{$-$}0.530 & \phantom{$-$}0.132\\
         & $C_{TSO}$ & $-$0.186 & $-$0.186 & $-$0.186\\
         & Total & \phantom{$-$}1.206 & \phantom{$-$}1.592 & \phantom{$-$}1.206\\
         & Excessive & \phantom{$-$}0.000 & \phantom{$-$}0.386 & \phantom{$-$}-\\
    \hline
         & $C_{Grid}$ & \phantom{$-$}0.630 & \phantom{$-$}0.615 & \phantom{$-$}0.635\\
         & $C_{DER}$ & \phantom{$-$}0.689 & \phantom{$-$}0.645 & \phantom{$-$}0.744\\
         & $C_{PV}$ & \phantom{$-$}0.000 & \phantom{$-$}0.000 & \phantom{$-$}0.000\\
        4 & $C_{Load}$ & \phantom{$-$}0.452 & \phantom{$-$}1.016 & \phantom{$-$}0.379\\
         & $C_{TSO}$ & $-$0.100 & $-$0.186 & $-$0.186\\
         & Total & \phantom{$-$}1.673 & \phantom{$-$}2.091 & \phantom{$-$}1.574\\
         & Excessive & \phantom{$-$}0.099 & \phantom{$-$}0.517 & \phantom{$-$}-\\
    \hline
         & $C_{Grid}$ & \phantom{$-$}0.619 & \phantom{$-$}0.610 & \phantom{$-$}0.619\\
         & $C_{DER}$ & \phantom{$-$}0.640 & \phantom{$-$}0.638 & \phantom{$-$}0.640\\
         & $C_{PV}$ & \phantom{$-$}0.000 & \phantom{$-$}0.000 & \phantom{$-$}0.000\\
        5 & $C_{Load}$ & \phantom{$-$}0.132 & \phantom{$-$}0.903 & \phantom{$-$}0.132\\
         & $C_{TSO}$ & $-$0.186 & $-$0.225 & $-$0.186\\
         & Total & \phantom{$-$}1.206 & \phantom{$-$}1.926 & \phantom{$-$}1.206\\
         & Excessive & \phantom{$-$}0.000 & \phantom{$-$}0.720 & \phantom{$-$}-\\
    \hline
    \end{tabular}
    %\vspace{-3mm}
    \label{t:costs}
\end{table}

The proposed method enables obtaining negligible excessive cost values for both cases with very small difference between the predicted and actual generation (case \#1), and cases with a significant value of this difference related to temporary shutdown of PV (case \#2, which can represent the occurrence of the PV unit fault) or the significantly lower generation than the prediction (case \#3, which can represent the occurrence of the unpredicted weather event). The reference deterministic model obtains similar results only for the case with accurate predictions, and relatively high (greater than 30\% of the optimal value) excessive costs for other cases.

In the proposed method, only PV generation is considered as stochastic, thus, the ability to deal with other types of unplanned events (such as unpredicted load demand increase -- case \#4, or unplanned flexibility service demand -- case \#5) is only a side effect of the stochastic approach to PV generation. Hence, the quality of the solution would depend significantly on the particular case. 
The cost distribution between different categories shows that, in general, the cost of the load shedding has the greatest impact on the cost differences between the proposed and reference method. In addition, for all cases the generation curtailment is avoided.
\section{Conclusion}\label{s:conclusion}
%The cost-effective and constraint-respecting integration of the renewable resources in the distribution system requires the appropriate distribution system control approaches.
This paper presents a decision-making method for the control of the distribution system with DERs equipped with PV and BESS. %This method can also be used by the DSO as an analytical instrument, e.g., to determine the optimal placement and parameters of DERs.
The objective of the approach is to minimize the total operational cost incurred by the DSO due to generation curtailment, load shedding, grid and DERs power losses, and taking into account the reward from the TSO for providing the flexibility service, while the system constraints are strictly followed. The transition probability of the PV generation enables utilization of the generation predictions from the previous day, and applying the same probability matrix for each time of the day and year. The results show that this method enables obtaining a relatively small value of the total cost, and the benefits are particularly evident when the difference between the actual and predicted value of PV generation is significant. Therefore, the application of such a method may potentially bring the greatest benefits in the case of networks particularly susceptible to extreme weather conditions or PV unit faults. Furthermore, the approach can also be used by DSOs as an analytical instrument, e.g., to determine the optimal placement and parameters of DERs. 
Since, the presented approach has computational overheads due to its calculation time that makes it not scalable for complex systems, our future work will  explore this aspect, e.g., the development of a distributed control strategy to minimize computational complexity.

\section*{Acknowledgments}
This publication is based upon
work supported by King Abdullah University of Science and Technology under Award No. ORFS-2022-CRG11-5021.

\bibliographystyle{ieeetr}
\bibliography{references}

\vfill

\end{document}